\documentclass[prl,aps,preprint,showpacs]{revtex4}
\usepackage{graphicx}

\begin{document}
\title{Influence of chemical pressure effects on nonlinear thermal conductivity of intrinsically granular superconductors}

\author{Sergei Sergeenkov}
\affiliation{Departamento de F\'isica, CCEN, Universidade Federal
da Para\'iba,\\ Cidade Universit\'aria, 58051-970 Jo\~ao Pessoa,
PB, Brazil}
\date{\today}

\begin{abstract}
Using a 2D model of capacitively coupled Josephson junction arrays
(created by a network of twin boundary dislocations with strain
fields acting as an insulating barrier between hole-rich domains
in underdoped crystals), we study the influence of chemical
pressure ($\nabla \mu$) on {\it nonlinear} (i.e., $\nabla T$ -
dependent) thermal conductivity (NLTC) of an intrinsically
granular superconductor. Quite a substantial enhancement of NLTC
is predicted when intrinsic chemoelectric field ${\bf E}_\mu
\propto \nabla \mu $ closely matches the externally produced
thermoelectric field ${\bf E}_T \propto \nabla T$. The estimates
of the model parameters suggest a realistic possibility to
experimentally monitor this effect in non-stoichiometric
high-$T_C$ superconductors.
\end{abstract}


\maketitle

High resolution imaging of the granular structure in underdoped
$Bi_2Sr_2CaCu_2O_{8+\delta}$ crystals~\cite{1} has recently
revealed an apparent segregation of its electronic structure into
superconducting domains (of the order of a few nanometers) located
in an electronically distinct background. In particular, it was
found that at low levels of hole doping ($\delta <0.2$), the holes
become concentrated at certain hole-rich domains. (In this regard,
it is interesting to mention a somewhat similar phenomenon of
"chemical localization" that takes place in materials, composed of
atoms of only metallic elements, exhibiting metal-insulator
transitions~\cite{2}.) Tunneling between such domains leads to
intrinsic granular superconductivity (GS) in high-$T_c$
superconductors (HTS). Probably one of the first examples of GS
was observed in $YBa_2Cu_3O_{7-\delta }$ single crystals in the
form of the so-called "fishtail" anomaly of
magnetization~\cite{3}. The granular behavior has been related to
the 2D clusters of oxygen defects forming twin boundaries (TBs) or
dislocation walls within $CuO$ plane that restrict supercurrent
flow and allow excess flux to enter the crystal. Indeed, there are
serious arguments to consider the TB in HTS as insulating regions
of the Josephson SIS-type structure.  Besides, a destruction of
bulk superconductivity in these non-stoichiometric materials with
increasing the oxygen deficiency parameter $\delta $ was found to
follow a classical percolation theory~\cite{4}. In addition to
their importance for understanding the underlying microscopic
mechanisms governing HTS materials, the above experiments can
provide rather versatile tools for designing chemically-controlled
atomic scale Josephson junctions (JJs) and their arrays (JJAs)
with pre-selected properties needed for manufacturing the modern
quantum devices~\cite{5,6}. To understand how GS manifests itself
in non-stoichiometric crystals, let us invoke an analogy with the
previously discussed dislocation models of grain-boundary
Josephson junctions (GBJJs) (see, e.g., \cite{7,8} and further
references therein). Recall that under plastic deformation, grain
boundaries (GBs) (which are the natural sources of weak links in
HTS), move rather rapidly via the movement of the grain boundary
dislocations (GBDs) comprising these GBs. At the same time,
observed~\cite{1,3,9,10,11} in HTS single crystals regular 2D
dislocation networks of oxygen depleted regions (generated by the
dissociation of $<110>$ twinning dislocations) with the size $d_0$
of a few Burgers vectors, forming a triangular lattice with a
spacing $d\ge d_0$ ranging from $10nm$ to $100nm$, can provide
quite a realistic possibility for existence of 2D Josephson
network within $CuO$ plane. Recall furthermore that in a $d$-wave
orthorhombic $YBCO$ crystal TBs are represented by tetragonal
regions (in which all dislocations are equally spaced by $d_0$ and
have the same Burgers vector ${\bf a}$ parallel to $y$-axis within
$CuO$ plane) which produce screened strain fields~\cite{10}
$\epsilon ({\bf x})=\epsilon (0)e^{-{\mid{{\bf x}}\mid}/d_0}$ with
${\mid{{\bf x}}\mid}=\sqrt{x^2+y^2}$. Though in
$YBa_2Cu_3O_{7-\delta }$ the ordinary oxygen diffusion
$D=D_0e^{-U_d/k_BT}$ is extremely slow even near $T_c$ (due to a
rather high value of the activation energy $U_d$ in these
materials, typically $U_d\simeq 1eV$), in underdoped crystals
(with oxygen-induced dislocations) there is a real possibility to
facilitate oxygen transport via the so-called osmotic (pumping)
mechanism~\cite{12,13} which relates a local value of the chemical
potential (chemical pressure) $\mu ({\bf x})=\mu (0)+\nabla \mu
\cdot {\bf x}$ with a local concentration of point defects as
follows $c({\bf x})=e^{-\mu ({\bf x})/k_BT}$. Indeed, when in such
a crystal there exists a nonequilibrium concentration of
vacancies, dislocation is moved for atomic distance $a$ by adding
excess vacancies to the extraplane edge. The produced work is
simply equal to the chemical potential of added vacancies. What is
important, this mechanism allows us to explicitly incorporate the
oxygen deficiency parameter $\delta $ into our model by relating
it to the excess oxygen concentration of vacancies $c_v\equiv
c(0)$ as follows $\delta=1-c_v$. As a result, the chemical
potential of the single vacancy reads $\mu _v\equiv \mu
(0)=-k_BT\log (1-\delta )\simeq k_BT\delta $. Remarkably, the same
osmotic mechanism was used by Gurevich and Pashitskii~\cite{10} to
discuss the modification of oxygen vacancies concentration in the
presence of the TB strain field. In particular, they argue that
the change of $\epsilon ({\bf x})$ under an applied or chemically
induced pressure results in a significant oxygen redistribution
producing a highly inhomogeneous filamentary structure of
oxygen-deficient nonsuperconducting regions along GB~\cite{11}
(for underdoped superconductors, the vacancies tend to concentrate
in the regions of compressed material). Hence, assuming the
following connection between the variation of mechanical and
chemical properties of planar defects, namely $\mu ({\bf
x})=K\Omega _0\epsilon ({\bf x})$ (where $\Omega _0$ is an
effective atomic volume of the vacancy and $K$ is the bulk elastic
modulus), we can study the properties of TB induced JJs under
intrinsic chemical pressure $\nabla \mu$ (created by the variation
of the oxygen doping parameter $\delta $). More specifically, a
single $SIS$ type junction (comprising a Josephson network) is
formed around TB due to a local depression of the superconducting
order parameter $\Delta ({\bf x})\propto \epsilon({\bf x})$ over
distance $d_0$ producing thus a weak link with (oxygen deficiency
$\delta $ dependent) Josephson coupling $J(\delta )=\epsilon({\bf
x})J_0=J_0(\delta )e^{-{\mid{{\bf x}}\mid}/d_0}$ where $J_0(\delta
)=\epsilon (0)J_0=(\mu _v/K\Omega _0 )J_0$. Thus, the considered
here model indeed describes chemically induced GS in underdoped
systems (with $\delta \neq 0$) because, in accordance with the
observations, for stoichiometric situation (when $\delta \simeq
0$), the Josephson coupling $J(\delta ) \simeq 0$ and the system
loses its explicitly granular signature.

There are several approaches for studying the thermal response of
JJs and JJAs  based on phenomenology of the Josephson effect in
the presence of thermal gradients (see, e.g.,~\cite{14,15,16,17}
and further references therein). In this paper, within a 2D model
of capacitive JJAs (created by a regular 2D network of TB
dislocations), we study the influence of chemical pressure effects
(described by the gradient of the chemical potential $\nabla \mu
$) on {\it nonlinear} (i.e. $\nabla T$-dependent) thermal
conductivity (NLTC) $\kappa (T,\nabla \mu ;\nabla T)$ of
intrinsically granular superconductors. As we shall see, in a
sharp contrast with its linear counterpart, NLTC exhibits quite a
pronounced enhancement when the intrinsically induced
chemoelectric field $E_\mu =\frac{1}{2e}|\nabla \mu |$ matches the
externally produced thermoelectric field $E_T=S_0|\nabla T|$.

To adequately describe transport properties of the chemically
induced granular superconductor under a simultaneous influence of
intrinsic chemical pressure $\nabla \mu ({\bf x})=K\Omega _0\nabla
\epsilon ({\bf x})$ and applied thermal gradient $\nabla T$, we
employ a model of 2D overdamped Josephson junction array which is
based on the following total Hamiltonian~\cite{12}
\begin{equation}
{\cal H}(t)={\cal H}_T(t)+{\cal H}_C(t),
\end{equation}
where
\begin{equation}
{\cal H}_T(t)=\sum_{ij}^NJ_{ij}[1-\cos \phi _{ij}(t)]
\end{equation}
is the well-known tunneling Hamiltonian, and
\begin{equation}
{\cal H}_C(t)=\sum_{ij}^N\frac{q_i(t)q_j(t)}{2C_{ij}}
\end{equation}
accounts for a Coulomb contribution due to mutual capacitances
$C_{ij}$ between grains with a net charge
\begin{equation}
q_{i}(t)=\sum_{j=1}^N\int_0^tdt'I_{ij}(t')
\end{equation}
induced by the normal current $I_{ij}(t)=V_{ij}(t)/R_{ij}$ within
the array, where $V_{ij}(t)=(\frac{\Phi _0}{2\pi})\frac{\partial
\phi _{ij}}{\partial t}$ and $R_{ij}$ are the corresponding
voltage and resistance between grains in their normal state.

According to the above-mentioned scenario, the tunneling
Hamiltonian ${\cal H}_T(t)$ introduces a short-range
(nearest-neighbor) interaction between $N$ junctions (which are
formed around oxygen-rich superconducting areas with phases $\phi
_i(t)$), arranged in a two-dimensional (2D) lattice with
coordinates ${\bf x_i}=(x_i,y_i)$. The areas are separated by
oxygen-poor insulating boundaries (created by TB strain fields
$\epsilon({\bf x}_{ij})$) producing a short-range Josephson
coupling $J_{ij}=J_0(\delta )e^{-{\mid{{\bf x}_{ij}}\mid}/d}$.
Besides, as we have seen, the strain fields $\epsilon({\bf
x}_{ij})$ also control the dielectric properties of the array.
Namely, the mutual capacitances between grains are defined as
$C_{ij}=2\pi \epsilon _0\epsilon({\bf x}_{ij})|{\bf
x}_{ij}|=C_0(|{\bf x}_{ij}|/d)e^{-{\mid{{\bf x}_{ij}}\mid}/d}$.
Thus, both the Josephson energy and the charging energy of the
array vary exponentially with the distance ${\bf x}_{ij}={\bf
x}_{i}-{\bf x}_{j}$ between neighboring junctions (with $d$ being
an average grain size). The temperature dependence of Josephson
coupling is governed by the standard expression~\cite{12}
\begin{equation}
J_{ij}(T)=J_{ij}(0)\left[\frac{\Delta (T)}{\Delta (0)}\right]\tanh
\left [\frac{\Delta (T)}{2k_BT}\right ]
\end{equation}
where $J_{ij}(0)=[\Delta (0)/2](R_0/R_{ij})$ with $\Delta (T)$
being the temperature dependent gap parameter; $R_0=h/4e^2$ and
$R_{ij}\propto J_{ij}^{-1}=R_{n}e^{{\mid{{\bf x}_{ij}}\mid}/d}$
stand for quantum and normal resistance, respectively. For the
explicit temperature dependence of the gap parameter we used the
analytical approximation~\cite{12}, namely $\Delta (T)=\Delta
(0)\tanh \left (2.2 \sqrt{\frac{T_C-T}{T}}\right )$ which is valid
for all temperatures.

By analogy with a constant electric field ${\bf E}$, a thermal
gradient $\nabla T$ applied to a chemically induced JJA will cause
a time evolution of the phase difference across insulating
barriers as follows~\cite{15,17,18}
\begin{equation}
\phi _{ij}(t)=\phi _{ij}^0+\frac{2e({\bf E}_\mu -{\bf E}_T){\bf
x}_{ij}}{\hbar}t
\end{equation}
Here $\phi _{ij}^0$ is the initial phase difference (see below),
${\bf E}_\mu =\frac{1}{2e}\nabla \mu$ and ${\bf E}_T=S_0\nabla T$
are the induced chemoelectric and thermoelectric fields,
respectively. $S_0$ is the Seebeck coefficient. In what follows,
we assume, for simplicity, that $\nabla \mu =(\nabla _x\mu ,0,0)$
with $\nabla _x\mu =\Delta \mu /d$, and that $\nabla T=(\nabla
_xT, 0,0)$.

In order to study the most interesting situation when the
externally produced thermoelectric field ${\bf E_T}$ becomes
comparable with intrinsically induced chemoelectric field ${\bf
E_\mu}$, let us consider a {\it nonlinear} generalization of the
conventional Fourier law and the resulting {\it nonlinear} thermal
conductivity (NLTC) under the influence of chemical pressure. In
what follows, by the NLTC we understand a $\nabla T$-dependent
thermal conductivity which is defined as follows~\cite{17}
\begin{equation}
\kappa (T,{\bf \nabla \mu}; \nabla T)\equiv -\frac{1}{V}\left
[\frac{\partial <Q_x>}{\partial (\nabla _{x}T)}\right ]_{\nabla
T\neq 0}
\end{equation}
where
\begin{equation}
<Q_{x}>=\frac{1}{\tau}\int_0^\tau dtQ_{x}(t)
\end{equation}
with $Q_x(t)$ being the longitudinal component of the total
thermal flux which is defined (in a q-space representation) via
the total energy conservation law as follows ($V$ is sample's
volume, and $\tau$ is a characteristic time of the problem, see
below)
\begin{equation}
{\bf Q}(t)\equiv \lim_{{\bf q} \to 0} \left [i\frac{{\bf q}}{{\bf
q}^2}{\dot{\cal H}_{\bf q}}(t)\right ]
\end{equation}
with
\begin{equation}
{\dot{\cal H}_{\bf q}}=\frac{1}{s}\int d^2x e^{i{\bf q}{\bf
x}}\frac{\partial {\cal H}({\bf x},t)}{\partial t},
\end{equation}
Here, $s=2\pi d^2$ is properly defined normalization area, and we
made a usual substitution~\cite{17} $\frac{1}{N}\sum_{ij}A_{ij}(t)
\to \frac{1}{s}\int d^2x A({\bf x},t)$ valid in the
long-wavelength approximation (${\bf q} \to 0$).

In view of Eqs.(1)-(10) and assuming $\phi _{ij}^0=\pi /2$ for the
initial phase difference (which maximizes the Josephson current of
the network), we arrive at the following analytical expression for
the temperature and chemical pressure dependence of the {\it
nonlinear} thermal conductivity (TC) of a model granular
superconductor
\begin{equation}
\kappa (T,{\bf \nabla \mu}; \nabla T)=\kappa _n\left [1+\beta
_C(T)\frac{(1+4\eta ^2)}{(1+\eta ^2)^4}\right ]
\end{equation}
Here, $\kappa _n\equiv \kappa (T_C,{\bf \nabla \mu}; \nabla
T)=2\pi Nd^2S_0I_C(0)/V\beta _C(0)$ is the normal state value of
the NLTC, $\beta _C(T)=2\pi I_C(T)C_{0}R_n^2/\Phi _0$ is the
so-called Stewart-McCumber parameter~\cite{19} responsible for
dissipative properties of the array with $I_c(T)=(2e/\hbar )J(T)$
being the critical current, and $\eta=(E_{\mu} -E_T)/E_0$ with
$E_0=\hbar /2ed\tau$ being a characteristic field. Notice that, as
expected, in the limit $E_T \to 0$ (or when $\nabla T =0$ in the
rhs of Eq.(7)), Eq.(11) reduces to the expression for the linear
thermal conductivity $\kappa (T,{\bf \nabla \mu}; 0)$.

\begin{figure}
\includegraphics[width=8cm]{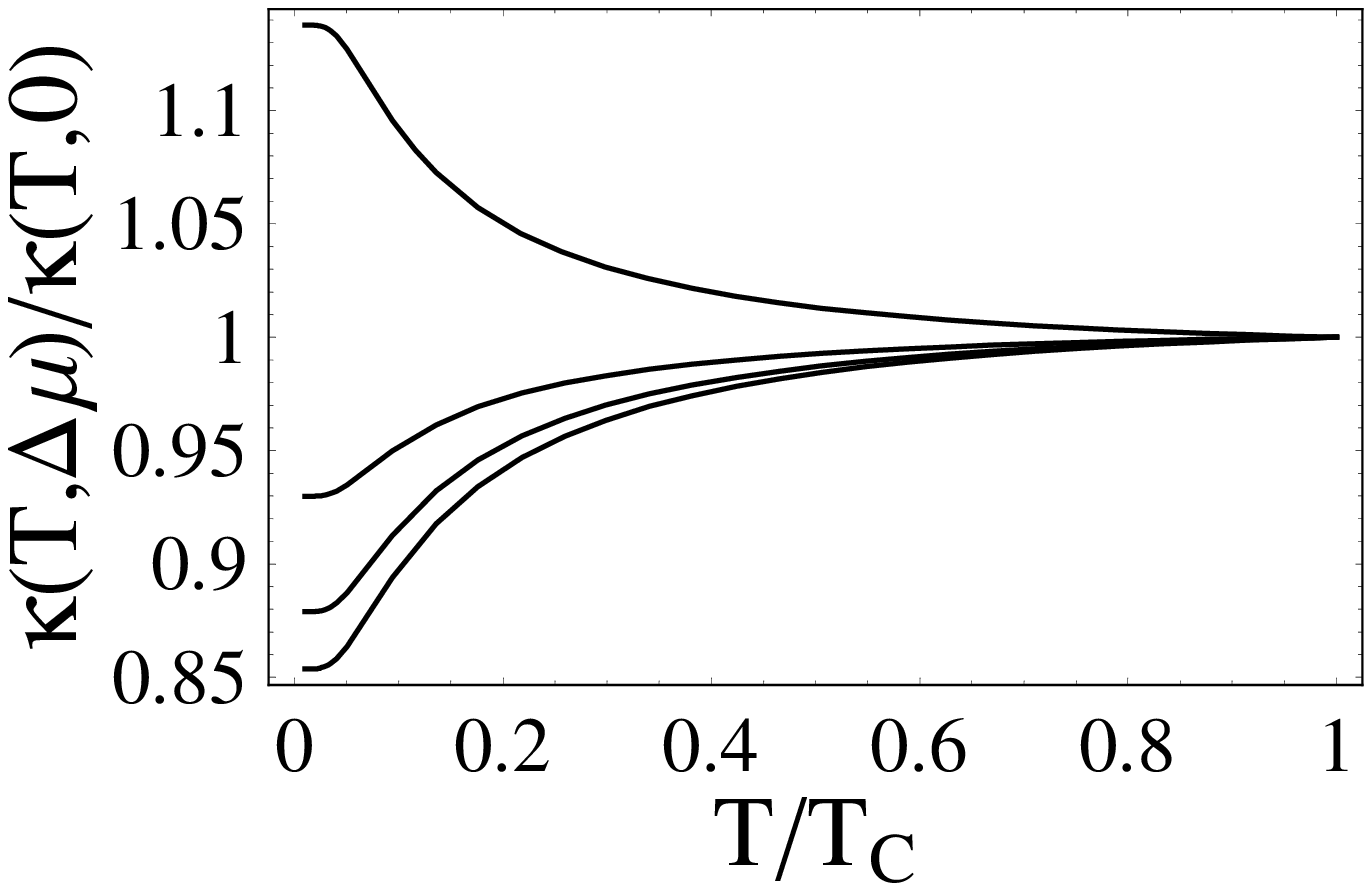}
 \caption{\label{fig:fig1} The  dependence of the  {\it nonlinear} thermal
 conductivity on reduced temperature $T/T_C$ for chemical pressure
  $\Delta \mu /\Delta \mu _0=1.0$ and for different values of the applied thermal
  gradient $\eta _T=S_0|\nabla T|/E_0$ ($\eta _T=0, 0.4, 0.8$, and $1.2$, increasing from top to bottom), according to Eq.(11). }
 \end{figure}

\begin{figure}
\includegraphics[width=8cm]{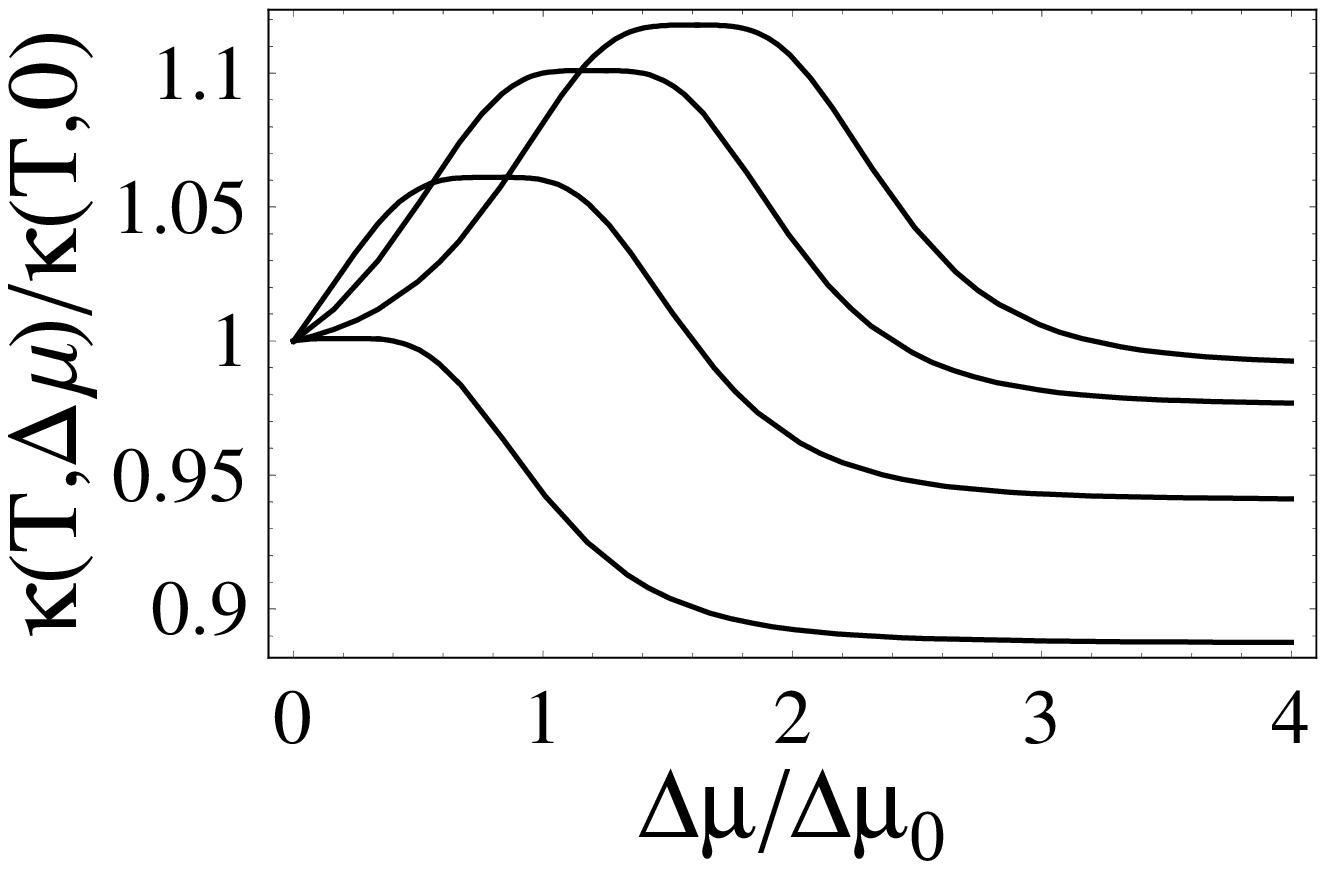}
 \caption{\label{fig:fig2} The dependence of the {\it nonlinear} thermal conductivity
  on the chemical pressure $\Delta \mu /\Delta \mu _0$
  for different  values of the applied thermal gradient $\eta _T=S_0|\nabla T|/E_0$
  ($\eta _T=0, 0.4, 0.8$, and $1.2$, increasing from bottom to top), according to Eq.(11). }
\end{figure}

Figure~\ref{fig:fig1} shows the  dependence of the normalized NLTC
$\kappa (T,\nabla \mu ; \nabla T)/\kappa (T,0; \nabla T)$ on
reduced temperature $T/T_C$ for chemical pressure $\Delta \mu
/\Delta \mu _0=1$ (with $\Delta \mu _0=\hbar /\tau$) for different
values of the applied thermal gradient $\eta _T=S_T|\nabla T|/E_0$
with $\beta _C(0)=1$ and $R_n=R_0$. Notice a markedly different
behavior of the nonlinear TC (corresponding to $\eta _T=0.4, 0.8,$
and $1.2$). Unlike its linear counterpart (shown by the upper
curve with $\eta _T=0$), it increases with increasing the
temperature. Even more drastic difference between the linear and
nonlinear TC can be seen in their chemical pressure dependence.
Figure~\ref{fig:fig2} depicts the behavior of the NLTC $\kappa
(T,\nabla \mu ; \nabla T)$ as a function of $\Delta \mu /\Delta
\mu _0$ for different values of the dimensionless parameter $\eta
_T=E_T/E_0$ and for $T=0.2T_C$ (the other parameters are the same
as before). As is clearly seen from this picture, in a sharp
contrast with the pressure behavior of the linear TC (shown by the
curve at the bottom and corresponding to $\eta _T=0$), its
nonlinear analog evolves with the chemoelectric field quite
differently. Namely, NLTC exhibits chemically stimulated
enhancement with a pronounced maximum at $E_\mu =\Delta \mu /2ed
\simeq E_T$. To complete our study, let us estimate the order of
magnitude of the main model parameters. Starting with
chemoelectric fields $E_\mu$ needed to observe the above-predicted
nonlinear field effects in granular superconductors, we notice
that according to Figure~\ref{fig:fig2}, the most interesting
behavior of NLTC takes place for $E_\mu \simeq E_0$. Using typical
$YBCO$ parameters~\cite{10}, $\epsilon (0)=0.01$, $\Omega
_0=a_0^3$ with $a_0=0.2nm$, and $K=115GPa$, we have $\mu
_v=\epsilon (0)K\Omega _0\simeq 1meV$ for an estimate of the
chemical potential in HTS crystals, which defines the
characteristic time $\tau \simeq \hbar /\mu _v \simeq 5\times
10^{-11}s$. Furthermore, taking $d\simeq 10nm$ for typical values
of the average grain size (created by oxygen-rich superconducting
regions), we get $E_0=\hbar /2ed\tau \simeq 5\times 10^{5}V/m$ and
$|\nabla \mu |=\Delta \mu /d \simeq \mu _v/d\simeq 10^{6}eV/m$ for
the estimates of the characteristic field and chemical potential
gradient (intrinsic chemical pressure), respectively. On the other
hand, the maximum of NLTC occurs when this field nearly perfectly
matches an "intrinsic" thermoelectric field $E_T=S_0|\nabla T|$
induced by an applied thermal gradient, that is when $E_\mu \simeq
E_0\simeq E_T$. Using~\cite{20} $S_0 \simeq 0.5\mu V/K$ for an
estimate of the linear Seebeck coefficient in $YBCO$, we obtain
$|\nabla T|\simeq E_0/S_0\simeq 2\times 10^6K/m$ for the
characteristic value of applied thermal gradient needed to observe
the predicted here effects. Let us estimate now the absolute value
of the normal state thermal conductivity $\kappa _n=2\pi
Nd^2S_0I_C(0)/V\beta _C(0)$. Recall that within the present
scenario, the scattering of normal electrons is governed by the
Stewart-McCumber parameter $\beta _C(T)=2\pi I_C(T)C_{0}R_n^2/\Phi
_0$ due to the presence of the normal resistance $R_n$ and mutual
capacitance $C_0$ between the adjacent grains. The latter is
estimated to be $C_0\simeq 1aF$ using $d=10nm$ for an average
"grain" size. Furthermore, the critical current $I_C(0)$ can be
estimated via the critical temperature $T_C$ as follows,
$I_C(0)\simeq 2\pi k_BT_C/\Phi _0$ which gives $I_C(0)\simeq 10\mu
A$ (for $T_C\simeq 90K$) and leads to $\beta _C(0)\simeq 3$ for
the value of the Stewart-McCumber parameter assuming $R_n\simeq
R_0$ for the normal resistance which, in turn, results in $q\simeq
\Phi _0/R_n\simeq 10^{-19}C$ and $E_C=q^2/2C_0\simeq 0.1eV$ for an
estimate of the "grain" charge and the Coulomb energy. Finally,
assuming $V\simeq Nd^3$ for the sample's volume, and using the
above-mentioned expressions for $S_0$ and $\beta _C(0)$, we obtain
$\kappa _n \simeq 10^{-3}W/mK$ for an estimate of the maximum of
the NLTC which is actually much higher than a similar estimate
obtained for inductance controlled $\kappa _n$ in electric-field
driven NLTC~\cite{17}, suggesting thus quite a realistic
possibility to observe the predicted here non-trivial behavior of
the thermal conductivity in non-stoichiometric high-$T_C$
superconductors.

This work was financially supported by the Brazilian agency CAPES.

\newpage

\newpage

\begin{thebibliography}{99}
\bibitem{1} K.M. Lang, V. Madhavan, J.E. Hoffman, E.W. Hudson, H. Eisaki, S. Uchida, and J. C. Davis, Nature {\bf 415},
 412 (2002).\\

\bibitem{2} V.F. Gantmakher, Physics-Uspekhi {\bf 45}, 1165
(2002); I.S. Beloborodov, A.V. Lopatin, V.M. Vinokur, and K.B. Efetov, Rev. Mod. Phys. {\bf 79}, 469 (2007).\\

\bibitem{3} M. Daeumling, J.M. Seuntjens, and D.C. Larbalestier, Nature
{\bf 346}, 332 (1990).\\

\bibitem{4} V.F. Gantmakher, A.M. Neminskii, and D.V. Shovkun,
JETP Lett. {\bf 52}, 630 (1990).\\

\bibitem{5} {\em New Directions in Mesoscopic Physics: Towards
Nanoscience}, Eds. R. Fazio, V.F. Gantmakher, and Y. Imry (Kluwer
Academic Publishers, Dordrecht, 2003).\\

\bibitem{6} I.V. Krive, S.I. Kulinich, R.I. Shekhter, and M. Jonson, Low Temp. Phys. {\bf 30}, 554
(2004).\\

\bibitem{7} E.Z. Meilikhov, JETP {\bf 83}, 803 (1996).\\

\bibitem{8} S. Sergeenkov, JETP Lett. {\bf 70}, 36 (1999).\\

\bibitem{9} G.Yang, P. Shang, S.D. Sutton, I.P. Jones, J.S. Abell, and C.E. Gough, Phys. Rev. B {\bf 48},
 4054 (1993).\\

\bibitem{10} A. Gurevich and E.A. Pashitskii, Phys. Rev. B {\bf 56},
 6213 (1997).\\

\bibitem{11} B.H. Moeckley, D.K. Lathrop, and R.A. Buhrman,
 Phys. Rev. B {\bf 47}, 400 (1993).\\

\bibitem{12} S. Sergeenkov, JETP Lett. {\bf 77}, 94 (2003); S. Sergeenkov, JETP {\bf 101}, 919 (2005);
S. Sergeenkov, in {\em Studies of High Temperature
Superconductors}, Ed. by  A.V. Narlikar (Nova Science Publishers,
New York, 2006), vol. {\bf 50}, p. 229.\\

\bibitem{13} S. Sergeenkov, J. Appl. Phys. {\bf 78}, 1114
(1995).\\

\bibitem{14} D. van Harlingen, D.F. Heidel, and J.C. Garland, Phys. Rev. B {\bf 21}, 1842
(1980).\\

\bibitem{15} G. Guttman, B. Nathanson, E. Ben-Jacob, and D.J. Bergman, Phys. Rev. B {\bf 55}, 12691
(1997).\\

\bibitem{16} J. Deppe and J.L. Feldman, Phys. Rev. B {\bf 50}, 6479
(1994).\\

\bibitem{17} S. Sergeenkov, JETP Lett. {\bf 76}, 170 (2002).\\


\bibitem{18} S. Sergeenkov, JETP Lett. {\bf 67}, 680 (1998).\\

\bibitem{19} A. Barone and G. Paterno, {\em Physics and Applications of the
Josephson Effect} (A Wiley-Interscience Publisher, New York,
1982).\\

\bibitem{20} A.V. Ustinov, M. Hartman, and R.P. Huebener,
Europhys. Lett. {\bf 13}, 175 (1990).


\end{thebibliography}
\end{document}